\documentclass[12pt]{article}%
\usepackage{citesort}
\usepackage{amsmath}
\usepackage{graphicx}%
\usepackage{amsfonts}%
\usepackage{amssymb}
\begin{document}

\title{Wounded quark-diquark model predictions\\for heavy ion collisions at the LHC}
\author{Adam Bzdak\thanks{e-mail: Adam.Bzdak@ifj.edu.pl}\\Institute of Nuclear Physics, Polish Academy of Sciences\\ul. Radzikowskiego 152, 31-342 Krakow, Poland}
\maketitle

\begin{abstract}
The ratios of particle densities in lead-lead and proton-lead collisions to
particle density in proton-proton collision in the central rapidity region at
the LHC energy are predicted on the basis of wounded quark-diquark model.

\vskip 0.6cm

\noindent PACS: 25.75.-q, 25.75.Ag, 21.65.Qr\newline Keywords: LHC, heavy ion
collisions, diquark, wounded nucleons

\end{abstract}

\section{Introduction}

The wounded quark-diquark model \cite{bb-AA1,bb-AA2} proved to be rather
successful in description of particles production from nuclear targets.
Assuming that high energy interactions of nucleons are dominated by
independent interactions of its two constituents, a quark and a diquark, it
was possible to describe $pp,$ $dAu,CuCu$ and $AuAu$ multiplicity data
collected at the RHIC collider \cite{phobos}. It indicates that in all
hadronic collisions the early stage of the particle production process can be
understood as a simple superposition of contributions from hadronic
constituents. As explained in \cite{bb-AA2} this does not preclude further
collective evolution of the system that is obviously present
\cite{v2-exp,wfwb}.

Encouraged by these results we present here quantitative predictions of the
wounded quark-diquark model for the particle density ratios $R_{AB}%
=N_{AB}/N_{pp}$ in the central rapidity region of $PbPb$ and $pPb$ collisions
at the LHC energy $\sqrt{s}=5500$ GeV.\footnote{The density of particles
produced in $pp$ collision at the LHC energy cannot be predicted in the
present approach.}

Our main conclusion is that the model provides rather precise predictions for
the nuclear collisions at LHC energies. This should allow its effective test
when the data are available.

In the next section the prediction of the wounded quark-diquark model for
particle density in the central rapidity region in $PbPb$ collision is
presented. In section $3$ we discuss the consequences of the model for
midrapidity density in $pPb$ collisions. Our conclusions are listed in the
last section where also some comments are included.

\section{PbPb collision}

The relation between particle production in nucleon-nucleon and symmetric
nucleus-nucleus collisions implied by the wounded quark-diquark model is given
by \cite{bb-AA1}%
\begin{equation}
R_{AA}\equiv\frac{N_{PbPb}(y)}{N_{pp}(y)}=\frac{w_{PbPb}^{(q+d)}}%
{2w_{p}^{(q+d)}},\label{RAA}%
\end{equation}
where the R.H.S of this equation is independent of rapidity $y$%
.\footnote{Provided we are far enough from the fragmentation regions, where
contributions from cascade and unwounded constituents are expected
\cite{bb-AA2}.} $N_{PbPb}(y)$ and $N_{pp}(y)$ are the particle densities in
$PbPb$ and $pp$ collisions, respectively. $w_{p}^{(q+d)}$ is the average
number of wounded constituents in a single $pp$ collision (per one proton).
Mean number of wounded quarks and diquarks in both colliding nuclei
$w_{PbPb}^{(q+d)}$ at a given impact parameter $b$ is given by (mass number
$A=208$) \cite{unm}%
\begin{equation}
w_{PbPb}^{(q+d)}(b)=\frac{2A}{\sigma_{PbPb}(b)}\int T(b-s)\left\{  2-\left[
1-p_{q}G(s)\right]  ^{A}-\left[  1-p_{d}G(s)\right]  ^{A}\right\}
d^{2}s,\label{wqd_pbpb}%
\end{equation}
with $G(s)$ defined as
\begin{equation}
G(s)=\int d^{2}s^{\prime}\sigma_{\text{in}}(s-s^{\prime})T(s^{\prime}),
\end{equation}
where $T(s)$ is the nuclear thickness function $T(s)=\int dz\rho(\sqrt
{s^{2}+z^{2}})$ (normalized to unity). Here and in the following for the
nuclear density $\rho$ we take the standard Woods-Saxon formula with the
nuclear radius $R_{Pb}=6.5$ fm and the skin depth $d=0.54$ fm. $\sigma
_{PbPb}(b)$ is the inelastic differential $PbPb$ cross
section.\footnote{$\sigma_{PbPb}(b)=1$, except at very large impact parameters
($b>14$ fm) which are of no interest.} Finally, $p_{q}$ and $p_{d}$ are the
probabilities for a quark and a diquark to interact in a single $pp$
collision, respectively.

We assume the differential inelastic $pp$ cross section $\sigma_{\text{in}%
}(s)$ (probability for inelastic $pp$ collision at a given impact parameter
$s$) to be in a simple Gaussian form\footnote{We believe that $\sigma
_{\text{in}}(0)=1$ is very close to reality. At ISR energies $\sigma
_{\text{in}}(0)=0.92$ \cite{zero}.}%
\begin{equation}
\sigma_{\text{in}}(s)=e^{-s^{2}/\varkappa^{2}},
\end{equation}
where $\varkappa^{2}=$ $\sigma_{\text{in}}/\pi$ and $\sigma_{\text{in}}$ is
the total inelastic $pp$ cross section $\sigma_{\text{in}}=%
{\textstyle\int}
\sigma_{\text{in}}(s)d^{2}s$.

The multiplicity data are usually presented versus the number of wounded
nucleons \cite{wnm}
\begin{equation}
w_{PbPb}^{(n)}(b)=\frac{2A}{\sigma_{PbPb}(b)}\int T(b-s)\left\{  1-\left[
1-G(s)\right]  ^{A}\right\}  d^{2}s. \label{wn_pbpb}%
\end{equation}
This completes all necessary formulas.

To obtain $w_{p}^{(q+d)}$ we followed exactly the procedure proposed at lower
energies, where we extracted this number \cite{bb-AA1} by studying
differential elastic $pp$ scattering cross section data. Indeed, assuming a
nucleon to be composed of a quark and a diquark, it was possible to describe
the small momentum transfer, $|t|<3$ GeV$^{2}$, elastic $pp$ and $\pi p$
scattering cross section data with a very high precision \cite{pp-pip}. In the
present case we studied the small $t$ elastic $p\bar{p}$ scattering data at
the Tevatron energy giving $w_{p}^{(q+d)}=1.24\pm0.01$. Considering many
different predictions regarding elastic $pp$ scattering at $14000$ GeV
\cite{pp-lhc} we obtained $w_{p}^{(q+d)}=1.28\pm0.02$. Thus in our
calculations at $\sqrt{s}=5500$ GeV for the average number of wounded quarks
and diquarks in a single $pp$ collision, per one colliding proton, we take
\begin{equation}
w_{p}^{(q+d)}=1.26\pm0.02.\label{wqd_p}%
\end{equation}

This number is the dominant uncertainty of our approach. The detailed
discussion of this problem, however, is beyond the scope of this investigation.

Since the total inelastic $pp$ cross section $\sigma_{\text{in}}$ is not known
at $\sqrt{s}=5500$ GeV we performed our calculations for three different
inelastic cross sections $\sigma_{\text{in}}=60$, $67$ and $75$ mb. We noticed
that at a given number of wounded nucleons we obtain practically the same
number of wounded constituents for each value of $\sigma_{\text{in}}$. This
observation allows for predictions at the LHC energy, which are practically
independent of the value of $\sigma_{\text{in}}$.

The calculated numbers for $\sigma_{\text{in}}=60$ mb are presented in Table
\ref{Tab_ppb}, where following \cite{pp-pip}, we assumed $p_{q}=p_{d}/2=$
$w_{p}^{(q+d)}/3$.\footnote{We also checked different choices, ranging from
$p_{d}=p_{q}$ to $p_{d}=2p_{q}$. We observe that the relation $w_{PbPb}%
^{(q+d)}$ vs $w_{PbPb}^{(n)}$ is not changed.} \begin{table}[h]
\begin{center}%
\begin{tabular}
[c]{|c|c|c|c|c|c|}\hline\hline
$%
\begin{array}
[c]{c}%
b\\
\text{\lbrack fm]}%
\end{array}
$ & $w_{PbPb}^{(n)}$ & $w_{PbPb}^{(q+d)}$ & $%
\begin{array}
[c]{c}%
b\\
\text{\lbrack fm]}%
\end{array}
$ & $w_{PbPb}^{(n)}$ & $w_{PbPb}^{(q+d)}$\\\hline\hline
0 & 409.2 & 793.3 & 8 & 179.9 & 315.2\\\hline
1 & 405.8 & 783.4 & 9 & 139 & 238.4\\\hline
2 & 394.3 & 753.5 & 10 & 101.6 & 169.6\\\hline
3 & 373.3 & 704.2 & 11 & 68.8 & 111.1\\\hline
4 & 343.2 & 638.9 & 12 & 42.1 & 65.2\\\hline
5 & 306.4 & 562.8 & 13 & 22.4 & 33.1\\\hline
6 & 265.4 & 480.7 & 14 & 9.9 & 14\\\hline
7 & 222.5 & 397 & 15 & 3.6 & 4.9\\\hline
\end{tabular}
\end{center}
\caption{Mean number of wounded quarks and diquarks $w_{PbPb}^{(q+d)}$ and
wounded nucleons $w_{PbPb}^{(n)}$ in $PbPb$ collision as a function of the
impact parameter $b$.}%
\label{Tab_pbpb}%
\end{table}

\begin{figure}[h]
\begin{center}
\includegraphics[scale=1.5]{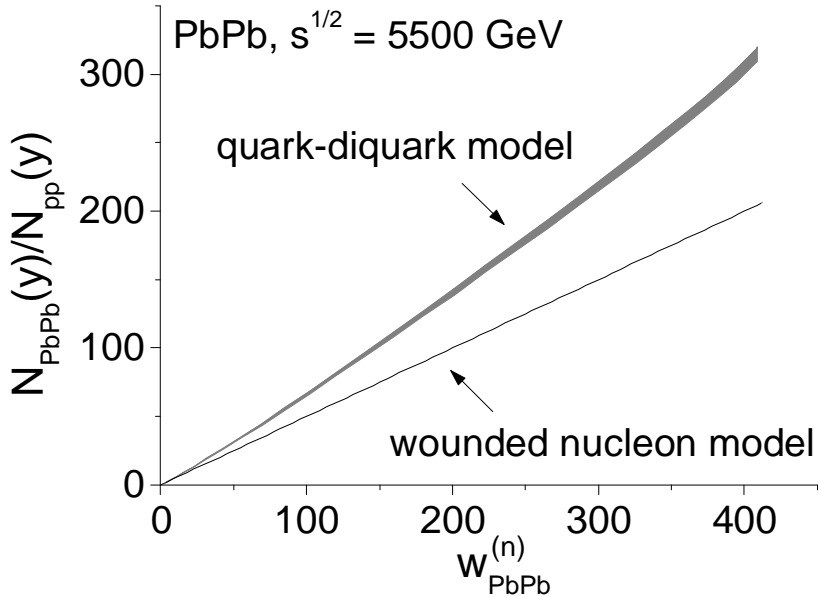}
\end{center}
\caption{{}Wounded quark-diquark model prediction for the multiplicity ratio
of particles produced in $PbPb$ collision to those produced in $pp$ collision
at any rapidity $y$. The grey band reflects the uncertainty in the value of
$w_{p}^{(q+d)}$. The prediction of the wounded nucleon model is also shown.}%
\label{fig_pbpb}%
\end{figure}

Dividing $w_{PbPb}^{(q+d)}$ by $2w_{p}^{(q+d)}$ we obtain our prediction for
the ratio $R_{AA}$ (\ref{RAA}) shown in Fig. \ref{fig_pbpb}. For comparison
the prediction of the wounded nucleon model \cite{wnm} is also
shown.

\section{pPb collision}

At the vanishing c.m. rapidity we have the following relation between particle
production in $pp$ and $pA$ collisions \cite{bb-AA1}
\begin{equation}
R_{pA}\equiv\frac{N_{pPb}(y=0)}{N_{pp}(y=0)}=\frac{w_{pPb}^{(q+d)}}%
{2w_{p}^{(q+d)}},\label{RpA}%
\end{equation}
where the average number of wounded quarks and diquarks in $pPb$ collision
$w_{pPb}^{(q+d)}$ at a fixed impact parameter $b$ is given by%
\begin{equation}
w_{pPb}^{(q+d)}(b)=\frac{AG(b)w_{p}^{(q+d)}}{1-\left[  1-G(b)\right]  ^{A}%
}+\frac{2-[1-p_{q}G(b)]^{A}-[1-p_{d}G(b)]^{A}}{1-\left[  1-G(b)\right]  ^{A}%
}.\label{wqd_ppb}%
\end{equation}

The first term gives the number of wounded constituents in the target ($Pb$
nucleus). Indeed, it is the number of wounded nucleons in the target times the
number of wounded constituents in a single $pp$ collision. The second term
gives the number of wounded constituents in the projectile that underwent many
inelastic collisions. The derivation of this term is presented in the appendix.

Mean number of wounded nucleons at a given impact parameter $b$ is given by%
\begin{equation}
w_{pPb}^{(n)}(b)=\frac{AG(b)}{1-\left[  1-G(b)\right]  ^{A}}+1, \label{wn_ppb}%
\end{equation}
where the first term gives the number of wounded nucleons in the target, plus
one wounded nucleon being the projectile itself.

Again, we performed the calculations for three different inelastic cross
sections $\sigma_{\text{in}}=60$, $67$ and $75$ mb. At a given impact
parameter $b$ we obtain significantly different numbers of wounded nucleons
and wounded constituents, however, when we plot $w_{pPb}^{(q+d)}$ vs
$w_{pPb}^{(n)}$ the three curves almost exactly follow each other. Similarly
to the previous case of $PbPb$ collision, this observation allows for
predictions at the LHC energy which are independent of the value of
$\sigma_{\text{in}}$.

The obtained numbers for $\sigma_{\text{in}}=75$ mb\footnote{This time we take
the largest number. Maximal number of wounded nucleons noticeably depends on
$\sigma_{\text{in}}$. The relation $w_{pPb}^{(q+d)}$ vs $w_{pPb}^{(n)}$ hardly
depends on it, however.} and $p_{q}=p_{d}/2=$ $w_{p}^{(q+d)}/3$ are presented
in Table \ref{Tab_ppb}. \begin{table}[h]
\begin{center}%
\begin{tabular}
[c]{|c|c|c|c|c|c|}\hline\hline
$%
\begin{array}
[c]{c}%
b\\
\text{\lbrack fm]}%
\end{array}
$ & $w_{PbPb}^{(n)}$ & $w_{PbPb}^{(q+d)}$ & $%
\begin{array}
[c]{c}%
b\\
\text{\lbrack fm]}%
\end{array}
$ & $w_{pPb}^{(n)}$ & $w_{pPb}^{(q+d)}$\\\hline\hline
0 & 17.05 & 22.23 & 6 & 6.78 & 9.2\\\hline
1 & 16.83 & 21.95 & 7 & 4.01 & 5.49\\\hline
2 & 16.15 & 21.08 & 8 & 2.6 & 3.47\\\hline
3 & 14.9 & 19.51 & 9 & 2.15 & 2.76\\\hline
4 & 12.92 & 17.01 & 10 & 2.03 & 2.57\\\hline
5 & 10.1 & 13.44 & 11 & 2.01 & 2.53\\\hline
\end{tabular}
\end{center}
\caption{Mean number of wounded quarks and diquarks $w_{pPb}^{(q+d)}$ and
wounded nucleons $w_{pPb}^{(n)}$ in $pPb$ collision as a function of the
impact parameter $b$.}%
\label{Tab_ppb}%
\end{table}

\begin{figure}[h]
\begin{center}
\includegraphics[scale=1.5]{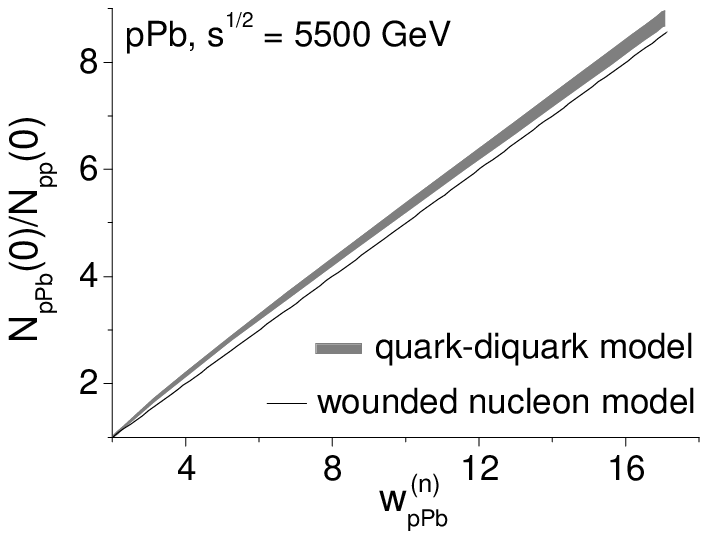}
\end{center}
\caption{Wounded quark-diquark model prediction for the multiplicity ratio at
midrapidity of particles produced in $pPb$ collision to those produced in $pp$
collision. The maximal number of wounded nucleons is $14$ and $17$ for
$\sigma_{\text{in}}=60$ and $75$ mb, respectively. The grey band reflects the
uncertainty in the value of $w_{p}^{(q+d)}$. The prediction of the wounded
nucleon model is also shown.}%
\label{fig_ppb}%
\end{figure}

Dividing $w_{pPb}^{(q+d)}$ by $2w_{p}^{(q+d)}$ we obtain our prediction for
the ratio $R_{pA}$ (\ref{RpA}) presented in Fig. \ref{fig_ppb}. The maximal
number of wounded nucleons is $14$ and $17$ for $\sigma_{\text{in}}=60$ and
$75$ mb, respectively. For comparison we also show the prediction of the
wounded nucleon model.

It is not surprising that the wounded quark-diquark model prediction is rather
close to the line predicted by the wounded nucleon model. Indeed, comparing
both scenarios the only difference is the projectile that undergoes many
inelastic collisions producing slightly more particles \cite{bb-AA2}.

\section{Conclusions and comments}

Our conclusions can be formulated as follows.

(i) Encouraged by a very good agreement of the wounded quark-diquark model
with the RHIC $pp,$ $dAu,CuCu$ and $AuAu$ data, we evaluated particle
densities in the central rapidity region in $PbPb$ and $pPb$ collisions at the
LHC energy $\sqrt{s}=5500$ GeV.

(ii) In our approach the particle density in $PbPb$ (at the central rapidity
region) and $pPb$ (at midrapidity) is proportional to the density of particles
produced in an elementary $pp$ collision. Since the $pp$ particle density is
presently unknown and it cannot be calculated in the present approach we only
give the ratios $R_{AB}=N_{AB}/N_{pp}$.

(iii) The dominant uncertainty of our calculation is the number of wounded
quarks and diquarks in a single $pp$ collision at $\sqrt{s}=5500$ GeV which we
estimated to be $1.26\pm0.02$.

(iv) Since the total inelastic $pp$ cross section is not known at $\sqrt
{s}=5500$ GeV we performed our calculations for three different inelastic
cross sections $\sigma_{\text{in}}=60$, $67$ and $75$ mb. The functional
relation between number of wounded quarks and diquarks and number of wounded
nucleons practically does not depend on the value of $\sigma_{\text{in}}$.
This observation allowed for predictions at the LHC energy, which are
independent of the value of $\sigma_{\text{in}}$.

Following comments are in order.

(a) Our prediction regarded the multiplicity density ratio $R_{AA}$ can be
also applied to the total multiplicities measured for central $PbPb$
collisions. For such centralities additional contributions from cascade and
unwounded constituents seem to be less important \cite{bb-AA2}.

(b) We found previously that the $200$ GeV RHIC data in the range $\left|
y\right|  <3.7$ can be solely described by the contribution from the wounded
constituents. Beyond this region unwounded constituents and cascade seems to
appear \cite{bb-AA2}. Assuming that these additional contributions begin at
$y$ proportional to the rapidity beam $Y$, it allows us to estimate that at
$\sqrt{s}=5500$ GeV the ratio $R_{AA}$ should be independent of $y$ in the
approximate range $\left|  y\right|  <6$.

(c) In principle our predictions could be applied to any energy provided
$\sigma_{\text{in}}$ remains in the range from $60$ mb to $75$ mb. The only
difference is the number of wounded quarks and diquarks in a single $pp$
collision. For instance at $\sqrt{s}=14000$ GeV we estimate $w_{p}^{(q+d)}$ to
be $1.28\pm0.02$.

\bigskip

\textbf{Acknowledgements}

We would like to thank A. Bialas for useful discussions and for critical
reading of the manuscript. This investigation was supported in part by the
Polish Ministry of Science and Higher Education, grant No. N202 034 32/0918.\ 

\appendix                                             

\section{Appendix: Wounded constituents in the projectile}

The average number of wounded quarks and diquarks in a nucleon that underwent
exactly $k$ inelastic collisions is given by%
\begin{equation}
w_{k}=1-(1-p_{q})^{k}+1-(1-p_{d})^{k},
\end{equation}
where $p_{q}$ and $p_{d}$ are the probabilities for a quark and a diquark to
interact in a single $pp$ collision, respectively.

The probability that the nucleon at a given impact parameter $b$ underwent
exactly $k$ inelastic collisions is given by a standard formula%
\begin{equation}
P_{k}(b)=\frac{1}{1-\left[  1-G(b)\right]  ^{A}}\dbinom{A}{k}\left[
G(b)\right]  ^{k}\left[  1-G(b)\right]  ^{A-k},
\end{equation}
where we assume that at least one inelastic collision takes place.

Thus, the number of wounded constituents in a nucleon that passed through the
nucleus of mass number $A$ at a given impact parameter $b$ is%
\begin{equation}
\sum\nolimits_{k=1}^{A}w_{k}P_{k}(b)=\frac{2-[1-p_{q}G(b)]^{A}-[1-p_{d}%
G(b)]^{A}}{1-\left[  1-G(b)\right]  ^{A}},
\end{equation}
$i.e.$ the second term in Eq. (\ref{wqd_ppb}).

\end{document}